\documentclass[aps,prappplied,reprint,showpacs,amsmath,amssymb,floatfix,superscriptaddress,noeprint]{revtex4-2}

\usepackage[normalem]{ulem}
\usepackage[colorlinks,linkcolor=blue,urlcolor=blue,citecolor=blue,anchorcolor=blue]{hyperref}

\usepackage{placeins}

\usepackage{multirow}
\usepackage{lipsum}

\usepackage{dsfont}

\usepackage{graphicx} 

\usepackage{physics}
\usepackage{siunitx}
\DeclareSIUnit{\belmilliwatt}{Bm}
\DeclareSIUnit{\dBm}{\deci\belmilliwatt}
\usepackage{tensor}

\usepackage{xcolor}

\usepackage{stackengine}
\stackMath

\makeatletter
\newsavebox\myboxA
\newsavebox\myboxB
\newlength\mylenA

\newcommand*\xoverline[2][0.9]{%
    \sbox{\myboxA}{$\m@th#2$}%
    \setbox\myboxB\null
    \ht\myboxB=\ht\myboxA%
    \dp\myboxB=\dp\myboxA%
    \wd\myboxB=#1\wd\myboxA
    \sbox\myboxB{$\m@th\overline{\copy\myboxB}$}
    \setlength\mylenA{\the\wd\myboxA}
    \addtolength\mylenA{-\the\wd\myboxB}%
    \ifdim\wd\myboxB<\wd\myboxA%
       \rlap{\hskip 0.5\mylenA\usebox\myboxB}{\usebox\myboxA}%
    \else
        \hskip -0.5\mylenA\rlap{\usebox\myboxA}{\hskip 0.5\mylenA\usebox\myboxB}%
    \fi}

\begin{document}
\title{Harnessing two-photon dissipation for enhanced quantum measurement and control}

\author{A. Marquet}
\affiliation{Alice \& Bob, 53 Bd du G\'en\'eral Martial Valin, 75015 Paris, France}
\affiliation{Ecole Normale Sup\'erieure de Lyon,  CNRS, Laboratoire de Physique, F-69342 Lyon, France}
\author{S. Dupouy}
\affiliation{Alice \& Bob, 53 Bd du G\'en\'eral Martial Valin, 75015 Paris, France}
\affiliation{Ecole Normale Sup\'erieure de Lyon,  CNRS, Laboratoire de Physique, F-69342 Lyon, France}
\author{U. R\'eglade}
\affiliation{Alice \& Bob, 53 Bd du G\'en\'eral Martial Valin, 75015 Paris, France}
\affiliation{Laboratoire de Physique de l’Ecole normale sup\'erieure, ENS-PSL, CNRS, Sorbonne Universit\'e, Universit\'e
Paris Cit\'e, Centre Automatique et Syst\`emes, Mines Paris, Universit\'e PSL, Inria, Paris, France}
\author{A. Essig}
\affiliation{Alice \& Bob, 53 Bd du G\'en\'eral Martial Valin, 75015 Paris, France}
\author{J. Cohen}
\affiliation{Alice \& Bob, 53 Bd du G\'en\'eral Martial Valin, 75015 Paris, France}
\author{E. Albertinale}
\affiliation{Alice \& Bob, 53 Bd du G\'en\'eral Martial Valin, 75015 Paris, France}
\author{A. Bienfait}
\affiliation{Ecole Normale Sup\'erieure de Lyon,  CNRS, Laboratoire de Physique, F-69342 Lyon, France}
\author{T. Peronnin}
\affiliation{Alice \& Bob, 53 Bd du G\'en\'eral Martial Valin, 75015 Paris, France}
\author{S. Jezouin}
\affiliation{Alice \& Bob, 53 Bd du G\'en\'eral Martial Valin, 75015 Paris, France}
\author{R. Lescanne}
\thanks{These authors co-supervised the project}
\affiliation{Alice \& Bob, 53 Bd du G\'en\'eral Martial Valin, 75015 Paris, France}
\author{B. Huard}
\thanks{These authors co-supervised the project}
\affiliation{Ecole Normale Sup\'erieure de Lyon,  CNRS, Laboratoire de Physique, F-69342 Lyon, France}

\date{\today}

\begin{abstract}
Dissipation engineering offers a powerful tool for quantum technologies. Recently, new superconducting devices demonstrated an engineered two-photon dissipation rate exceeding all other relevant timescales. In particular, they have proven most useful to prevent transitions between the logical states $|\pm\alpha\rangle$ of a cat qubit. Here, we present three key applications of strong two-photon dissipation for quantum measurement and control, beyond cat qubit stabilization. Firstly, we demonstrate its efficacy in overcoming limitations encountered in Wigner tomography at high photon numbers. Secondly, we showcase its potential for realizing universal gates on cat qubits, exploiting the coherent mapping between cat qubit states and superpositions of 0 and 1 photons. Finally, we harness the transient dynamics of a cat state under two-photon dissipation to prepare squeezed cat states with a squeezing factor exceeding $3.96 \pm 0.07$ dB.
\end{abstract}

\maketitle

\section{Introduction}

The perception of dissipation in quantum systems has undergone a drastic change. Once seen as a relentless foe, it is now being recognized as a potential ally, capable of being `tamed' to become a valuable tool for quantum measurements, control, and quantum error correction~\cite{Zoller1996, Beige2000, Harrington2022, Guillaud2023, Miao2023}.
Within the framework of bosonic codes, dissipation offers a strategy to remove entropy from a \emph{memory} mode (e.g., superconducting resonator, trapped ions, spin ensemble) and thereby stabilize a logical qubit~\cite{Ralph2003, Jeong2007, Mirrahimi2014, Leghtas853, Albert2019, Touzard2018, Lescanne2020exponential, Gertler2021, deNeeve2022, Gautier2022, Berdou2022, Xu2023, Reglade2023, Marquet2023, Sivak2023, lachancequirion2023}. This is achieved by carefully tailoring the coupling between the memory and its environment, inducing an effective dissipation that confines information within the desired qubit subspace and prevents leakage into the broader Hilbert space. This technique has proven particularly successful in stabilizing \emph{cat qubits},~\cite{Mirrahimi2014, Leghtas853, Touzard2018, Lescanne2020exponential, Berdou2022, Reglade2023, Marquet2023}, whose computational basis comprises two coherent states $\ket{\pm \alpha}$. A key resource in this process is two-photon dissipation, where the memory loses pairs of photons to a low-temperature environment at a rate $\kappa_2$. A two-photon drive acting on the memory completes the stabilization process of the cat qubit.

This work explores the broader potential of two-photon dissipation beyond cat qubit stabilization. We leverage the \emph{autoparametric cat} device~\cite{Marquet2023}, which boasts an engineered two-photon dissipation rate $\kappa_2$ exceeding the residual single-photon loss rate $\kappa_1 =(11~\mu\mathrm{s})^{-1}$ by two orders of magnitude. Following proposals in~\cite{Albert2016,Chamberland2022}, we harness two-photon dissipation to extend Wigner tomography protocols to large photon numbers, and achieve universal control of cat qubits within a timescale $\kappa_2^{-1}$. Furthermore, we demonstrate the preparation of squeezed cat states with a squeezing parameter reaching 3.9 dB. 

\begin{figure}[h!]
\includegraphics[width=\linewidth]{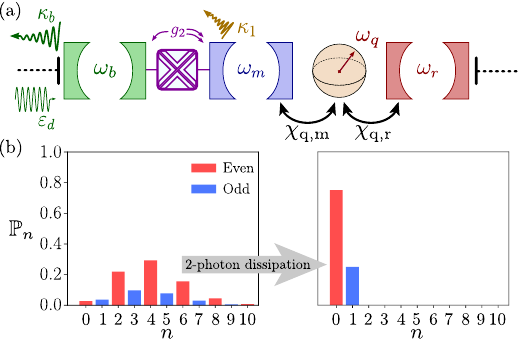}
\caption{(a) Schematic of a typical device for two-photon dissipation. The memory mode (blue) interacts with the buffer mode (green) via a 2-to-1 photon exchange with coupling strength $g_2$. The buffer can be driven with an amplitude $\epsilon_d$. Additionally, a dispersively coupled tomography transmon (orange) with frequency $\omega_q$  can be read out through a dedicated readout resonator at $\omega_r$. (b) Photon number distribution $\mathbb{P}_n$ of an arbitrary quantum state before and after undergoing two-photon dissipation.}
\label{fig:fig1}
\end{figure}

To implement such a two-photon dissipation, a dedicated mode, denoted as the buffer, is used to mediate the interaction between the memory and its environment~\cite{Leghtas853, Touzard2018, Lescanne2020exponential, Berdou2022, Reglade2023,Marquet2023}. In our experiment, this buffer dissipates photons at a rate $\kappa_b/2\pi\approx 40~\mathrm{MHz}$ (Fig.~\ref{fig:fig1}a). The memory and buffer modes are coupled through a nonlinear element implementing a two-to-one photon exchange Hamiltonian $\hat{H}=\hbar g_2 \hat{m}^2 \hat{b}^\dagger +h.c.$ at a rate $g_2$. Here, $\hat{m}$ and $\hat{b}$ represent the annihilation operators of the memory and buffer modes, respectively. This interaction is realized in our experiment using a 3-wave mixing interaction under the condition $2 \omega_m = \omega_b$, characteristic of autoparametric systems~\cite{Marquet2023} (see Sec.~\ref{Sec:Autoparametric_coupling_analogy}).

Operating in the regime $\kappa_b \geq 8 g_2 |\alpha|$ ensures that photons in the buffer mode are evacuated into the environment at a faster rate than they are converted into pairs of photons in the memory. This allows for the adiabatic elimination of the buffer mode, leading to an effective dissipation operator $\sqrt{\kappa_2} \hat{m}^2$ where $\kappa_2 = 4 g_2^2 / \kappa_b$.  In the autoparametric cat device~\cite{Marquet2023}, an external flux bias $\phi_\mathrm{ext}$ can be used to activate or deactivate two-photon dissipation. We have identified a specific flux point, $\phi_\mathrm{on}$, at which  $\omega_b=2\omega_m$, resulting in  $\kappa_2/2\pi = 2.16~\mathrm{MHz}$. Another flux point, $\phi_\mathrm{off}$ with negligible $\kappa_2/2\pi$ serves to disable two-photon dissipation (details in Sec.~\ref{Sec:standard_Wigner_measurement}). It is chosen at a point where the self-Kerr effect is weak.

\section{Dissipation enhanced Wigner tomography}

Let us begin by demonstrating how two-photon dissipation can address limitations in Wigner tomography, as proposed in~\cite{Chamberland2022}. The Wigner function of a density matrix $\hat{\rho}$ is proportional to the expectation value of the parity operator $\hat{\mathcal{P}} = e^{i \pi \hat{m}^\dagger \hat{m}}$after the memory has undergone a displacement operation $\hat{\mathcal{D}} \left(- \beta \right) = e^{-(\beta \hat{m}^\dagger - \beta^* \hat{m})}$:  
\begin{equation}
\label{eq:expression_Wigner_parity}
     W_{\hat{\rho}}\left(\beta \right) = \frac{2}{\pi} \mathrm{Tr}\left( \hat{\mathcal{D}}\left(\beta \right)^\dagger \hat{\rho} \,\hat{\mathcal{D}}\left(\beta \right) \mathcal{\hat{P}}\right).
\end{equation}
The displacement is achieved by driving the memory on resonance with a pulse whose complex amplitude scales with $-\beta$. In our system, the pulse duration of $20~\mathrm{ns}$ is chosen to be significantly shorter than the inverse self-Kerr rate of the memory $\chi_\mathrm{m, m}\approx (720~\mathrm{ns})^{-1}$. Wigner tomography then boils down to measuring the average value of $\hat{\mathcal{P}}$. 

Various parity measurement methods have been demonstrated in superconducting circuits, usually relying on coupling the memory at resonance \cite{Hofheinz2008, Hofheinz2009} or dispersively~\cite{Lutterbach1997,Bertet2002,Haroche2006,Kirchmair2013,Vlastakis2013} to an auxiliary qubit.  However, all these methods deteriorate as the average photon number $\bar{n}$ after displacement increases. Let us focus on the Ramsey-like sequence proposed in Ref.~\cite{Lutterbach1997}. The qubit is first prepared in an equal superposition of ground and excited state $(|g\rangle+|e\rangle)/\sqrt{2}$ by applying an unconditional $\pi/2$ pulse, much shorter than the inverse dispersive coupling, which is  $(\chi_\mathrm{q,m})^{-1}\approx0.9~\mu\mathrm{s}$ in our experiment. After a time $T_\mathrm{parity} = \pi / \chi_\mathrm{q, m}$, the superposition accumulates a phase $n \pi$ with $n$ the number of photons of the memory. Even and odd numbers of photons are then respectively mapped onto the states $\left(\ket{g} + \ket{e}\right) / \sqrt{2}$ and $\left(\ket{g} - \ket{e}\right) / \sqrt{2}$  of the qubit. A final unconditional $\pi/2$ pulse is applied, mapping the memory parity operator $\hat{\mathcal{P}}$ onto the Pauli operator $\hat{\sigma}_z=|e\rangle\langle e|-|g\rangle\langle g|$ of the qubit, which can then be readout. 

Firstly, this protocol underestimates the amplitude of the Wigner function since the parity outcome is flipped when a photon is lost during the idle time $T_\mathrm{parity}$, which occurs with probability $1-e^{-\kappa_1T_\mathrm{parity}\bar{n}}$. Additionally,  for large  $\bar{n}$, the superconducting qubit state can leak out of its subspace $\mathrm{span}(|g\rangle,|e\rangle)$, especially after the qubit gets excited with the $\pi/2$-pulse~\cite{Sank2016, Lescanne2019, Shillito_2022, Cohen2023,Xiao2023, dumas_2024}. Finally, the upper bound on the $\pi/2$ pulse duration scales as $(\chi_\mathrm{q, m}\bar{n})^{-1}$, requiring impractically high driving power at large photon numbers. These limitations at high photon numbers can be overcome using the protocol proposed in Ref.~\cite{Chamberland2022}. The key lies in leveraging two-photon dissipation's ability to preserve parity while transforming a state with an arbitrary photon number distribution into a state with at most $1$ photon (Fig.~\ref{fig:fig1}b).

\begin{figure}[h!]
\includegraphics[width=\linewidth]{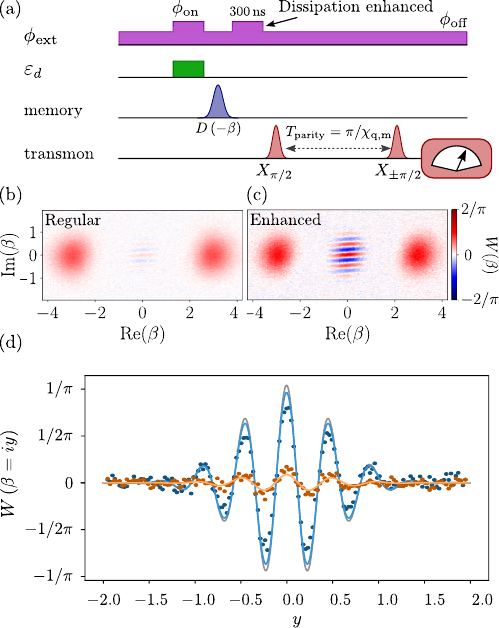}
\caption{(a) Pulse sequence of the dissipation enhanced Wigner tomography of $\ket{C_\alpha^+}$ including an extra 300~ns idle time at $\phi_\mathrm{ON}$ after memory displacement. (b) Left: Wigner tomography of $\ket{C_\alpha^+}$, $\alpha = 3.3$, without the extra idle time. (c) Wigner tomography of $\ket{C_\alpha^+}$ using the dissipation enhancement. (d) Cuts of the Wigner function along the imaginary axis $\mathrm{Re(\beta)} = 0$ in various scenarios. Solid gray line: simulated memory state at the end of the 300~ns buffer pulse $\epsilon_d$. Orange points: Wigner tomography in (b). Solid orange line: simulation of the measurement protocol without the extra idle time. Blue points: Wigner tomography in (c). Solid blue line: simulation of the dissipation enhanced measurement protocol (see Sec.~\ref{sec:parity_meas}).}
\label{fig:fig2}
\end{figure}

To illustrate typical errors of Wigner tomography, we prepare a cat state $\ket{C_\alpha^+}\propto|\alpha\rangle+|-\alpha\rangle$ and realize its tomography. The preparation is performed by tuning the flux to $\phi_\mathrm{on}$, which turns on the two-photon dissipation, and driving the buffer mode on resonance for 300~ns with a drive amplitude $\epsilon_d  = -g_2^* \alpha^2$.  As demonstrated in Refs.~\cite{Leghtas853,Marquet2023}, this implements an effective dissipation operator $\hat{L}_2 = \sqrt{\kappa_2}\left(\hat{m}^2 - \alpha^2 \right)$. Fig.~\ref{fig:fig2}b shows the obtained Wigner tomography for $\alpha=3.3$. The typical fringes of cat states can be seen close to $\beta=0$ but their contrast is much smaller than expected for $\ket{C_\alpha^+}$. Part of this loss of contrast comes from single photon losses during preparation. However, in the cut $W_{\hat{\rho}}(\beta)$ along $\beta\in i\mathbb{R}$ (Fig.~\ref{fig:fig2}.d), one can see that the measured Wigner function (orange dots) does not faithfully reproduce numerical simulations only accounting for the preparation infidelity (grey line).  
Here, most of the tomography errors can be attributed to single photon loss during parity measurement. Indeed, the probability of losing a photon during $T_\mathrm{parity}\approx 2.7~\mu\mathrm{s}$  reads $ 1-e^{-0.26\bar{n}}$ when the memory has $\bar{n}$ photons on average, reaching $50~\%$ for 2.7 photons. The error is therefore dramatic for the cat state in Fig.~\ref{fig:fig2}.b which has more than 10 photons on average. This effect is reproduced by simulations (orange line in  Fig.~\ref{fig:fig2}.d). In contrast, the measurement of $W_{\hat{\rho}}(\pm\alpha)$ is only marginally affected by single photon loss. Indeed, after a displacement $\hat{\mathcal{D}} \left(\pm \alpha \right)$, the memory ends up in a superposition of $|0\rangle$ and $|\pm 2 \alpha \rangle$. The parity of  $|0\rangle$ does not depart from 1 during measurement and the parity of $|\pm 2 \alpha \rangle$ is negligible, leading to a measured $W_{\hat{\rho}}(\pm\alpha)\approx 1/\pi$ as expected (see Fig.~\ref{fig:fig2supmat}).

We solve this issue by quickly turning the two-photon dissipation back on by tuning the flux to $\phi_\mathrm{on}$ after the displacement (Fig.~\ref{fig:fig2}a). Idling at $\phi_\mathrm{on}$ for $300~\mathrm{ns}$ (a few $\kappa_2^{-1}$) then ensures that the memory decays to a state with at most 1 photon while preserving its parity (Fig.~\ref{fig:fig1}b). Tuning the flux back to $\phi_\mathrm{off}$ enables parity measurement. In our example of a cat state above, a better measurement of the Wigner fringes can then be achieved (Fig.~\ref{fig:fig2}c,d). Simulations corresponding to this parity measurement protocol (blue line) predict a drastic improvement of the fringe reconstruction, reproducing up to $5~\%$ what is expected when only accounting for preparation infidelity (grey line). This is confirmed by the measured Wigner function (blue dots) which shows an increase in contrast compared to the usual parity measurement protocol. It should be noted that avoiding single-photon loss while removing pairs of photons requires a larger $\kappa_2/\kappa_1$ ratio. This condition can be met with other devices than the autoparametric cat, such as a simple parametrically pumped transmon qubit~\cite{Touzard2018}, making this method accessible to a wide range of experiments using superconducting qubits.  Note that we do not find an upper bound in photon number for this tomography technique and manage to demonstrate it up to 25 photons. We note that this protocol was used in Ref.~\cite{Marquet2023}.

\section{Quantum gates on cat qubits}

\begin{figure}[ht!]
\centering
\includegraphics[width = \linewidth]{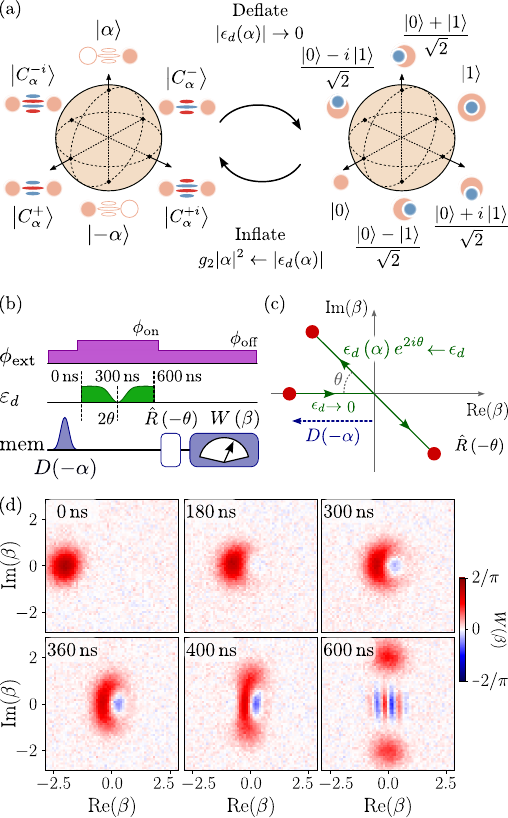}
\caption{(a) Inflation and deflation processes coherently map the manifolds $\mathrm{span}\left(\ket{\alpha}, \ket{-\alpha} \right)$ and $\mathrm{span}\left(\ket{0}, \ket{1} \right)$ by turning on or off the buffer drive $\epsilon_d$~\cite{Albert2016,Reglade2023}. (b) Pulse sequence used to prepare an initial state $\ket{\alpha}$, apply the holonomic $\hat{X}\left(\theta \right)$ gate, and measure the memory Wigner function. (c) Schematic evolution in the memory phase space of a coherent state while applying the holonomic $\hat{X}\left(\theta \right)$ gate. (d) Measured evolution of the memory Wigner function, starting from the coherent state $\ket{-2.27}$, during the $\hat{X}\left(\pi/2 \right)$ gate. The last snapshot occurs right before the final virtual rotation $\hat{R}(-\pi/2)$. Every measured time can be found in the appendix Video~\ref{vid:Xdynamics}.}
\label{fig:fig3}
\end{figure}

Two-photon dissipation can also be used to control cat qubits. For instance, driving the memory on resonance while the dissipation $\hat{L}_2$ is turned on performs a gate $\hat{Z}(\theta)$~\cite{Touzard2018,Marquet2023,Reglade2023} . In contrast, $\hat{X}(\theta)$ and $\hat{Y}(\theta)$ gates are seemingly incompatible with the engineered dissipation $\hat{L}_2$, which prevents population transfer between $|\alpha\rangle$ and $|-\alpha\rangle$. However, one can use the coherent mapping between the manifolds $\mathrm{span}\left(\ket{0}, \ket{1} \right)$ and $\mathrm{span}\left(\ket{\alpha}, \ket{-\alpha} \right)$ to perform these gates while the memory transits through the $\{|0\rangle,|1\rangle\}$ manifold (Fig.~\ref{fig:fig3}a)~\cite{Albert2016}. These \emph{deflation} and \emph{inflation} processes are performed by adiabatically modulating the buffer drive between $|\epsilon_d | = 0$ and $|\epsilon_d | = g_2 \left|\alpha \right|^2$, so that the memory follows a steady state of the dissipation. From this point of view, the preparation of $\ket{C_\alpha^+}$ in Fig.~\ref{fig:fig2} corresponds to the inflation of the vacuum state, and the Wigner tomography is enhanced by deflation before parity measurement. 

We implement a \textit{holonomic} $\hat{X}\left(\theta \right)$ gate by modulating the buffer drive amplitude $\epsilon_d(\alpha)$ while remaining at $\phi_\mathrm{on}$ (Fig.~\ref{fig:fig3}b). We first turn off the buffer drive to deflate the memory state into the $\{|0\rangle,|1\rangle\}$ manifold. The buffer drive is then turned back on with an additional phase $2\theta$, $\epsilon_d \rightarrow \epsilon_d \left( \alpha \right) e^{2 i \theta}$ (see Sec.~\ref{Sec:Holo_gate_appendix}), mapping the memory into the $\{|\alpha e^{i\theta}\rangle,|-\alpha e^{i\theta}\rangle\}$ manifold. This additional phase implements a virtual $\hat{X}(\theta)$ gate in the $\{|0\rangle,|1\rangle\}$ manifold  (see Fig.~\ref{fig:fig3}a). A rotation in the memory phase space $\hat{R} \left(-\theta \right)$, which maps any coherent state $|\beta e^{i\theta}\rangle$ onto $|\beta\rangle$, is finally applied to bring the memory back into the cat code while preserving the phase of the superposition. This final rotation could be performed by adiabatically sweeping the phase of the buffer drive back to $0$. Equivalently, we choose to apply an instantaneous virtual operation by changing the phase of the local oscillator of the memory drive by $-\theta$. The evolution of the memory in phase space starting from $\ket{-\alpha}$ is represented in Fig.~\ref{fig:fig3}c. 
We probe the evolution of the memory during the $\hat{X}\left(\pi/2 \right)$ gate starting from the coherent state $\ket{-\alpha}$ with $\alpha = 2.27$. This particular gate has been used for tomography without a transmon in~\cite{Reglade2023}. Wigner tomographies of the memory are shown in Fig.~\ref{fig:fig3}d at $6$ different times. The memory state can first be seen evolving from the coherent state towards $\left(\ket{0} - \ket{1} \right)/\sqrt{2}$  (third snapshot) before being mapped into $\left(\ket{\alpha e^{i \pi/2}} - i\ket{-\alpha e^{i \pi/2}} \right)/\sqrt{2}$ (last snapshot). 

We show in Fig.~\ref{fig:X_Holo_trace_distance} and Video~\ref{vid:Xgatedynamics} the state of the memory after applying the $\hat{X}\left(\theta \right)$ gate on the same coherent state $\ket{-\alpha}$ for an arbitrary angle $\theta$. Note that the shape of the buffer drive during the gate was optimized with numerical simulations shown in Fig.~\ref{fig:X_gate_opti}. The gate error being substantial, we settle for characterizing the gate properties by its effect on the coherent state $|-\alpha\rangle$ only. We compute the trace distance $T\left(\hat{\rho}, |\psi(\theta)\rangle\langle\psi(\theta)|\right)$~\cite{Nielsen2000} (see Eq.~(\ref{eq:tracedistance})) between the density matrix $\hat{\rho}$ reconstructed from the measured Wigner function and the state $|\psi(\theta)\rangle = \hat{X}\left(\theta \right)|-\alpha\rangle$. The largest distance occurs at $\theta=\pm \pi/2$, where it reaches $0.23$ (see Fig.~\ref{fig:X_Holo_trace_distance}). Numerical simulations qualitatively reproduce the dependence of the trace distance as a function of $\theta$. Larger trace distances are obtained for angles close to $\theta = \pm \pi/2$, corresponding to states exhibiting coherence, owing to the impact of phase flip errors during the inflation and stabilization step of the gate. Additionally, deviation between the simulations and measurements can be observed at those angles, which we explain by measurement infidelity (deviations between blue dots and grey line in Fig.~\ref{fig:fig2}d). Note that measurement infidelity is larger for cat states than for coherent states. Finally, simulations with and without the dephasing rate $\kappa_\varphi^\mathrm{m}/2\pi = 0.08~$MHz reveal it has a major contribution to the gate error owing to induced decoherence when the memory is in the $\{|0\rangle,|1\rangle\}$ manifold. The remaining gate errors observed when $\kappa_\varphi^\mathrm{m}/2\pi = 0.00~$MHz (blue line in Fig.~\ref{fig:X_Holo_trace_distance}b) can be attributed to a non-adiabatic evolution of the memory during the gate, or single photon loss.

\section{Preparing squeezed cat states}

\begin{figure}[ht!]
\centering
\includegraphics[width = \linewidth]{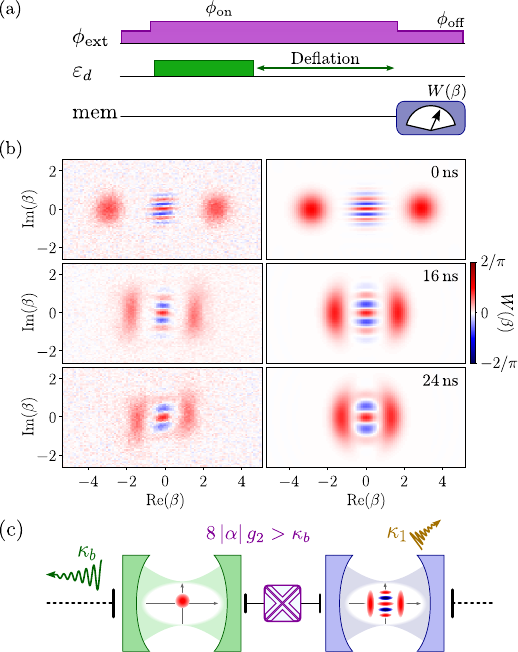}
\caption{Preparation of a squeezed cat state. (a) Pulse sequence. (b) Wigner functions of the memory at early times (0, 16~ns, or 24~ns) during the deflation of $\ket{C_\alpha^+}$, where $\alpha=2.9$. The measurements (left) are compared with numerical simulations made without free parameters (right). (c) Schematic representation of the buffer and memory states, in the non-adiabatic regime, at the beginning of deflation. An effective displacement of mode $\hat{b}$ creates an effective squeezing Hamiltonian acting on the memory during its decay to the $\left\{\ket{0}, \ket{1} \right\}$ manifold.} 
\label{fig:fig4}
\end{figure}

Interestingly, probing the deflation dynamics at short times (a few $\kappa_b^{-1}$) reveals a potential resource for state preparation. The deflation spontaneously prepares squeezed cat states.
After preparing the state $\ket{C_\alpha^+}$ in the memory, the drive $\epsilon_d (\alpha)$ is turned off while the flux remains at $\phi_\mathrm{on}$ (Fig.~\ref{fig:fig4}a). The measured Wigner functions of the memory state are shown in Fig.~\ref{fig:fig4}b for three different times between 0 and 24~ns. 
As the memory releases photon pairs to the buffer, it acts as an effective drive on the buffer with an amplitude $\left<g_2 \hat{m}^2 \right> = g_2 \alpha^2$. This drive, which should not have an impact on the buffer when $\kappa_b \gg 8 g_2 \alpha$, here displaces the buffer into a state whose mean amplitude $\langle \hat{b} \rangle$ is proportional to $\alpha^2$. In turn an effective squeezing Hamiltonian 

\begin{equation}
    \hat{H}_\mathrm{m, eff}/ \hbar = g_2 \langle \hat{b} \rangle \hat{m}^{\dagger 2} + h.c
\end{equation}

\noindent acts on the memory, which compresses the superposition of coherent states (Fig.~\ref{fig:fig4}c). As the memory deflates, it reaches a maximal squeezing at a time that decreases with the initial photon number. This observation is compatible with the expected evolution of the squeezing parameter $r$ at small time in $(g_2|\alpha| t)^2$ (see Sec.~\ref{Sec:squeezing}). From there, the buffer decays to vacuum, hence suppressing the effective squeezing Hamiltonian. To extract the squeezing ratio $e^{2r}$, we fit the two measured peaks in the Wigner function with Gaussian functions of identical variance. We observe squeezing ratios as high as $e^{2r}\approx3.96 \pm 0.07~\mathrm{dB}$~\cite{lvovsky2016}, obtained after a time $t = 16$~ns, starting from $\ket{C_\alpha^+}$ with $\alpha = 2.9$ (Fig.~\ref{fig:fig4}b). Note that during the evolution of the memory, the product of the largest and smallest variances remains close to that of the vacuum state. The predicted evolution using the bipartite buffer-memory system, without the adiabatic elimination, reproduces qualitatively the measured Wigner functions (right panels). A rounding of the Wigner function appears in the theory owing to two-photon dissipation rate, which ultimately overcomes the squeezing rate as the buffer is depleted. While bigger compression is expected starting with larger values of $\left |\alpha \right|$, an uncontrolled deformation of the initial cat state in the experiment limits the squeezing factor. We attribute this deformation and the discrepancy between simulations and experiments to the self-Kerr, the dephasing of the buffer or a detuning between $2\omega_m$ and $\omega_b$.

\section{Conclusion}

This work illustrates how two-photon dissipation is a remarkable tool for quantum measurement and control. Let us put in perspective the three applications we demonstrated. The enhancement of Wigner tomography is straightforward for any device with $\kappa_2\gg\kappa_1$ and should be useful for a broad range of experiments. Furthermore, the demonstration of an arbitrary $\hat{X}(\theta)$ gate together with the known $\hat{Z}(\theta)$ gate~\cite{Touzard2018,Marquet2023,Reglade2023} provides universal control of cat qubits. We note that since $\hat{X}(\pi)$ gate can be performed virtually by changing the phase of the local oscillator of the memory, such a control is not required for quantum error correcting codes~\cite{Guillaud2019}. However, our experiment in Fig.~\ref{fig:fig3} provides a useful tool to explore and optimize the dynamics of the holonomic gate, which is an essential component of cat qubit readout without transmon qubits~\cite{Reglade2023}. Finally, two-photon dissipation appears to be an interesting approach for the preparation of squeezed cat states, whose coherence properties are promising for quantum information processing~\cite{LeJeannic2018,Schlegel2022,Pan2023,Xu2023,Hillmann2023}.  

\begin{acknowledgments} 
This research was supported by the QuantERA grant QuCos ANR-19-QUAN-0006, the Plan France 2030 through the projects ANR-22-PETQ-0003 and ANR-22-PETQ-0006. We acknowledge IARPA and Lincoln Labs for providing a Josephson Traveling-Wave Parametric Amplifier. We thank Zaki Leghtas and Mazyar Mirrahimi for inspiring discussions and feedback.
\end{acknowledgments}

\appendix

\section{Mechanical analogy for the autoparametric coupling}
\label{Sec:Autoparametric_coupling_analogy}

\begin{figure}[h!]
\includegraphics[scale = 1.1]{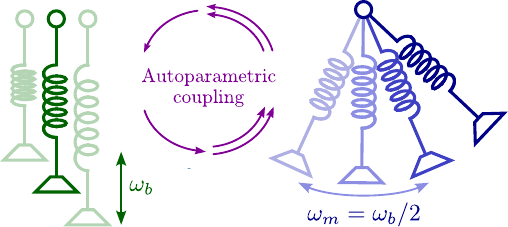}
\caption{Mechanical analogy illustrating the concept of autoparametric coupling. A mass $m$ is fixed to a spring with constant $k$ and equilibrium length $l$, connected on its other end to a rotation axis.}
\label{fig:Autoparametric_coupling}
\end{figure}

The principle of autoparametric coupling can be intuited using a mechanical analogy\cite{Tondl2000}. Consider a mass $m$ suspended on a spring of constant $k$, itself fixed on its other end to a rotation axis (Fig.~\ref{fig:Autoparametric_coupling}). The spring length is $l$ at rest. All other masses of the system are considered negligible compared to $m$. This system presents two distinct mechanical modes of motion, which can be thought of as the buffer and memory modes in our experiment. First, an oscillation corresponding to successive compression and stretching of the spring at a characteristic frequency $\omega_b=\sqrt{k/m}$ (green in the figure). Second, a pendulum-like oscillation where the spring rotates around the rotation axis at a frequency $\omega_m=\sqrt{g/l}$ (blue in the figure), with $g$ the Earth's gravity. Under the frequency matching condition $\sqrt{k/m} = 2 \sqrt{g/l}$, an energy transfer spontaneously occurs between these two modes. An initial elongation of the spring (buffer) is then converted into an oscillation around the rotation axis (memory), itself converted into compression-stretching oscillations, and so on.  

\section{Measurement of the Wigner function}
\subsection{Details on the experimental implementation of parity measurement}
\label{Sec:standard_Wigner_measurement}

In this section, we provide additional details about the parity measurement scheme of Fig.~\ref{fig:fig2}.

A first subtlety, which is common practice, consists of alternating the rotation angle of the last unconditional pulse in the Ramsey-like sequence between $\pi/2$ and $-\pi/2$. The signals corresponding to positive and negative final $\pi/2$ pulses are subtracted and renormalized to obtain $W\left(\beta \right)$. Note that this subtraction is theoretically unnecessary as this measurement could be performed using successive $\pi/2$ pulses only. However, doing so allows for the removal of unwanted offsets that could originate from the experimental setup. Additionally, it allows to correct for measurement errors introduced by the inefficiency of  $\pi/2$ pulses at large photon numbers (for pulse durations longer than $(n\chi_\mathrm{q,m})^{-1}$, the $\pi/2$ gate is inneffective). Indeed for large Fock states $|n\rangle$, the transmon remains close to the ground state during the $\pi/2$ pulses so that it is more and more likely to readout the qubit in the ground state as $n$ increases. The subtraction of the signals coming from the interleaved $\pi/2$ and $-\pi/2$ pulse sequences then suppress this effect. 

\label{sec_04_Wigner_fast_flux_line}

Secondly, we point out that this measurement protocol would be ineffective without tuning the flux away from $\phi_\mathrm{on}$ due to the strong two-photon dissipation. Indeed, the engineered dissipation impedes proper displacements $\hat{D}\left(-\beta \right)$, the very first step of the pulse sequence shown in Fig.~\ref{fig:fig2}a,  if the drive amplitude is smaller or comparable to the rate $\kappa_2$. Furthermore, it would also render impossible the parity measurement using the transmon qubit by inhibiting the dispersive coupling. One way to grasp the origin of the inhibition is to picture that engineered dissipation broadens the linewidth of the memory energy levels by $\kappa_2$. The qubit transition frequencies for all photon numbers then overlap in the regime $\kappa_2 \gg \chi_\mathrm{q,m}$ that is reached at $\phi_\mathrm{on}$. Ultimately, this prevents the qubit from having the desired dynamics during the idle step between the $\pi/2$ pulses of the protocol.

To overcome these challenges, the Wigner tomography is performed at $\phi_\mathrm{off}$ where $\left| \omega_{b, \mathrm{tomo}} - 2\omega_{m, \mathrm{tomo}} \right|/2\pi > 1~\mathrm{GHz}$. At this flux, two-photon dissipation is disabled because the 2-photon exchange Hamiltonian is not preserved under the rotating wave approximation. To rapidly switch between $\phi_\mathrm{on}$ and $\phi_\mathrm{off}$, we make use of a fast flux line that is designed to set the desired magnetic flux in approximately $20~\mathrm{ns}$.

While the memory dynamics at $\phi_\mathrm{off}$ is primarily dominated by the self-Kerr rate $\chi_\mathrm{m,m}/2\pi \approx 220~\mathrm{kHz}$, which only marginally affects the state during the 20~ns it takes to switch the flux, it is crucial to keep the drive $\epsilon_d\left(\alpha \right)$ on while shifting from $\phi_\mathrm{on}$ to $\phi_\mathrm{off}$. The memory dynamics at $\phi_\mathrm{on}$ is indeed dominated by the 2-photon dissipation with a rate $\kappa_2$, which significantly impacts the system in 20~ns, as demonstrated in Fig.~\ref{fig:fig4}. To prevent state distortion before the Wigner tomography, the drive $\epsilon_d\left(\alpha \right)$ is thus maintained during the flux change. This drive at $\omega_{b, \mathrm{2 ph}}$ does not affect the memory at $\phi_\mathrm{off}$, where the frequency matching condition is no longer satisfied.

\begin{figure}[h!]
\includegraphics[width=\columnwidth]{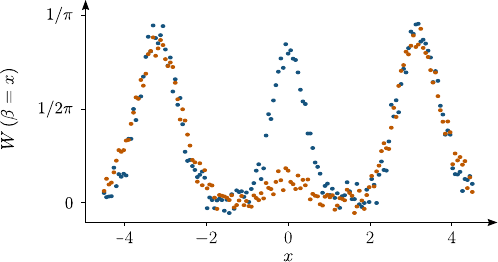}
\caption{Cuts of the Wigner function shown in Fig.~\ref{fig:fig2}b (orange) and Fig.~\ref{fig:fig2}c (blue) along the real axis $\mathrm{Im(\beta)} = 0$.}
\label{fig:fig2supmat}
\end{figure}

Figure \ref{fig:fig2supmat} shows a horizontal cut of the Wigner functions in Fig.~\ref{fig:fig2}.

\subsection{QNDness of the Wigner tomography}

In this brief subsection, we argue that our parity measurement method can be made Quantum Non Demolition in the context of a cat code. A QND measurement of parity consists of starting from an arbitrary state $|\Psi \rangle$ and measuring the observable $\hat{\Pi}=\hat{P}_\mathrm{even}-\hat{P}_\mathrm{odd}$. Here, $\hat{P}_\mathrm{even}=\sum_n|2n\rangle\langle 2n|$ and $\hat{P}_\mathrm{odd}=1-\hat{P}_\mathrm{even}$. If the outcome $1$ (resp. $-1$) is found, the state becomes $\hat{P}_\mathrm{even} |\Psi \rangle / \sqrt{\langle \Psi | \hat{P}_\mathrm{even} | \Psi \rangle}$ (resp. $\hat{P}_\mathrm{odd} |\Psi \rangle / \sqrt{\langle \Psi | \hat{P}_\mathrm{odd} | \Psi \rangle}$). In our procedure, however, once the parity measurement has been performed, the memory is always projected on either the Fock state $|0\rangle$ or $|1\rangle$ depending on the measurement outcome. This method, while allowing a better measurement of the parity, is then non QND in the general context of bosonic modes. 

For cat qubits where, starting from a state $|\Psi \rangle =a|C_\alpha^+ \rangle + b|C_\alpha^- \rangle$ (with $|a|^2 + |b|^2 = 1$), a QND measurement would project the memory into the cat states $|C_\alpha^+ \rangle$ or $|C_\alpha^-\rangle$. For this encoding, one could realize an inflation step (turning on the buffer drive) after the memory is projected onto the Fock state $|0\rangle$ or $|1\rangle$, to effectively collapse the memory state onto $|C_\alpha^+\rangle$ or $|C_\alpha^-\rangle$, making the parity measurement QND. Such a technique could theoretically be possible for other bosonic qubits, mapping $|0\rangle$ or $|1\rangle$ to the corresponding states $\hat{P}_\mathrm{even} |\Psi \rangle / \sqrt{\langle \Psi | \hat{P}_\mathrm{even} | \Psi \rangle}$ or  $\hat{P}_\mathrm{odd} |\Psi \rangle / \sqrt{\langle \Psi | \hat{P}_\mathrm{odd} | \Psi \rangle}$ in the encoding. However, it might not be as easily performed as for cat qubits where a single additional pump is required after the parity measurement. 

\subsection{Simulations of the parity measurement}
\label{sec:parity_meas}

This section descibes the model we use for the numerical simulations presented as solid lines in Fig.~\ref{fig:fig2}d. We simulate our system using the master equation
\begin{equation}
\begin{aligned}
\label{Eq:simu_parity_fidelity}
    \frac{d\hat \rho}{dt} =& -\frac{i}{\hbar}[\hat{H} , \hat \rho] + \kappa_1\mathcal{D}[\hat{m} ](\hat \rho)  + \kappa_2\mathcal{D}[\left( \hat{m}^2 - \alpha^2 \right)](\hat \rho)\\
    &+ \Gamma_\uparrow\mathcal{D}[\sigma_+](\hat \rho) +  \Gamma_\downarrow\mathcal{D}[\sigma_-](\hat \rho) + \frac{\Gamma_\varphi}{2}\mathcal{D}[\sigma_z](\hat \rho),
\end{aligned}
\end{equation}
where $\Gamma_\uparrow+\Gamma_\downarrow=1/T_1$, $\Gamma_\uparrow/\Gamma_\downarrow=(1+1/n_{th, q})^{-1}$ and the dephasing rate is $\Gamma_\varphi=1/T_2 - 1/2T_1$. The time-independent Hamiltonian is expressed as
\begin{equation}
\begin{aligned}
\label{Eq:simu_parity_fidelity_Hamiltonian}
    \frac{\hat{H}}{\hbar} = \,  - \chi_\mathrm{q,m} \hat{m}^\dagger \hat{m} ~\sigma_z.
\end{aligned}
\end{equation}

\noindent The ladder operators $\sigma_+$, $\sigma_-$ and the Pauli operator $\sigma_z$ act on the transmon, here approximated as a qubit. In these equations, $\kappa_1$ and $\kappa_2$ correspond to the single and two-photon loss rate of the memory, $T_1$ and $T_2$ are the energy decay time and coherence time of the transmon qubit, $n_{th, q}$ is the thermal population of the transmon, and $\chi_{q, m}$ is the cross-Kerr between the transmon and memory. The value of these parameters can be found in Table.~\ref{Tab:Device_parameters}. Because we consider different regimes of fluxes and buffer drive during the parity measurement, the values of $\kappa_2$ and $\alpha$ can be tuned to match the state of the system at each stage of the parity measurement. 

We first simulate the solid grey line in Fig.~\ref{fig:fig2}d corresponding to an imperfect cat preparation followed by an ideal parity measurement. The preparation is simulated using the master equation Eq.~(\ref{Eq:simu_parity_fidelity}) with $\kappa_2/2\pi = 2.16~\mathrm{MHz}$ and $\alpha = 3.3$, inflating the memory from a vacuum state during 300~ns. We then compute the ideal Wigner function of the obtained state, which shows a large deviation from $|C_\alpha^+ \rangle$ (for which $W(0) = 2/\pi$) that we attribute to single photon loss during the inflation. 

We then compare this ideal case to what would be measured using the usual parity measurement protocol, the solid orange line in Fig.~\ref{fig:fig2}d. During this measurement, the external flux is set to $\phi_\mathrm{off}$ which we take into account by setting $\kappa_2 = 0~\mathrm{MHz}$. Starting from the imperfectly prepared state previously computed, we first apply a displacement $\hat{\mathcal{D}}\left(-\beta \right)$ on the memory to simulate the measured value of the Wigner at point $\beta$.  We then perform an ideal $\pi/2$ pulse on the transmon (we neglect the error of this pulse compared to preparation errors or errors occurring during the idling time), after which the system is left idle for a time $\pi/\chi_{q, m}$. Finally, a second ideal $\pi/2$ is applied on the qubit, after which the parity can be deduced from the qubit state, reproducing the measurement of Fig.~\ref{fig:fig2}c.

Finally, we consider the case where pairs of photons are removed prior to the parity measurement (solid blue line in Fig.~\ref{fig:fig2}d). We once again start from the imperfectly prepared state after the inflation process and apply a displacement $\hat{\mathcal{D}}\left(-\beta \right)$. To simulate the deflation, we then remain idle at $\phi_\mathrm{on}$ for 300~ns, during which $\kappa_2/2\pi = 2.16~\mathrm{MHz}$ and $\alpha$ is set to $0$ in Eq.~(\ref{Eq:simu_parity_fidelity}). Note that in our simulations, we adiabatically eliminate the buffer and only considered errors originating from single photon loss or transmon errors during the deflation. This could in particular explain the discrepancy between this simulation and the experimental data shown in Fig.~\ref{fig:fig2}d (blue dots). Once the deflation toward the manifold $\mathrm{span}\left(|0\rangle, |1 \rangle \right)$ is applied, we simulate the usual parity measurement sequence with two ideal $\pi / 2$ pulses interleaved by an idling time $\pi/\chi_{q, m}$.

\section{Performing the gates $\hat{X}(\theta)$ and $\hat{Y}(\theta)$ using the manifold $\mathrm{span}\left(\ket{0}, \ket{1} \right)$}

\subsection{Performing the $\hat{Y}\left(\theta \right)$ gate}

While the existence of $\hat{Z}(\theta)$ and $\hat{X}(\theta)$ gates with arbitrary angles are sufficient for universal control, we describe in this section how one may perform a $\hat{Y}(\theta)$ gate directly~\cite{Iyama2024}.

\begin{figure}[h!]
\centering
\includegraphics[scale = 1.1]{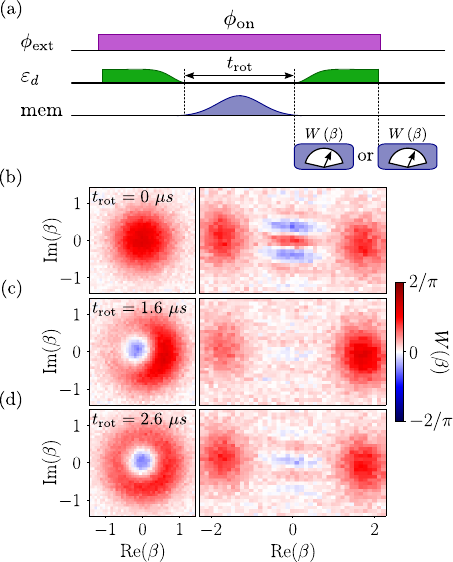}
\caption{(a) Pulse sequence for the gate $\hat{Y}\left(\theta \right)$. First, the buffer drive is turned off with the same pulse shape as for the $\hat{X}(\theta)$ gate, hence mapping the memory state onto the deflated manifold. The memory is then driven resonantly at $\omega_m$  with a Gaussian amplitude $\epsilon_\mathrm{Y}(t)$ for a time $t_\mathrm{rot}$. Finally, the memory is mapped back onto the cat encoding by turning the buffer drive back on. Wigner tomography is performed after the memory drive or at the end of the pulse sequence. (b-d) Measured Wigner functions of the memory after driving times $t_\mathrm{rot}=0$ (b), $1.6~\mu$s (c), or $2.6~\mu$s  (d). Corresponding inflated cat qubit states after the buffer drive are shown on the right panels. Video~\ref{vid:Ygate} shows every measured duration.}
\label{fig:Evol_X_gate_Zeno_blocked}
\end{figure}

We start by deflating the cat qubit state  into the manifold $\mathrm{span}\left(\ket{0}, \ket{1} \right)$. One may then perform a rotation around any axis on this qubit by driving the memory close to resonance, as in any two-level system. We implement the gate $\hat{Y}\left(\theta \right)$ on the cat qubit by driving the memory on resonance at $\phi_\mathrm{on}$ with an amplitude $\epsilon_\mathrm{Y}\in\mathbb{R}$ (Fig.~\ref{fig:Evol_X_gate_Zeno_blocked}a). Owing to quantum Zeno dynamics~\cite{Bretheau2015}, the combined effects of two-photon dissipation $\sqrt{\kappa_2} \hat{m}^2$ and buffer drive keep the memory state into the qubit space generated by the Fock states $\{0\rangle,|1\rangle\}$. The only constraint is to keep $|\epsilon_Y|\ll \kappa_2$ to meet the criterion of quantum Zeno dynamics.  The buffer drive is then turned back on, which maps the memory back onto the cat manifold. 

The measured Wigner functions of the memory starting in $\ket{C_\alpha^+}$, after the rotation or after the inflation, are shown in Fig.~\ref{fig:Evol_X_gate_Zeno_blocked}b-d and Video~\ref{vid:Ygate}. for a fixed displacement amplitude $\epsilon_\mathrm{Y}$ and three durations $t_\mathrm{rot}$. When no displacement is applied ($t_\mathrm{rot}=0$), the memory remains in $\ket{0}$ which is then mapped to $\ket{C_\alpha^+}$ after turning the buffer drive back on (Fig.~\ref{fig:Evol_X_gate_Zeno_blocked}b). When applying the displacement drive for $1.6~\mu$s and $2.6~\mu$s, the memory evolves towards $0.57\ket{0} + 0.82\ket{1}$ and $\ket{1}$ respectively (Fig.~\ref{fig:Evol_X_gate_Zeno_blocked}c,d). These states are then mapped onto the cat manifold to states resembling $0.57\ket{C_\alpha^+} + 0.82\ket{C_\alpha^-}$ and $\ket{C_\alpha^-}$. However, a noticeable deviation from these ideal states can be seen after the gate, particularly visible in the poorly contrasted fringes of what should be $\ket{C_\alpha^-}$.

\begin{video}
  \includegraphics[width=\columnwidth]{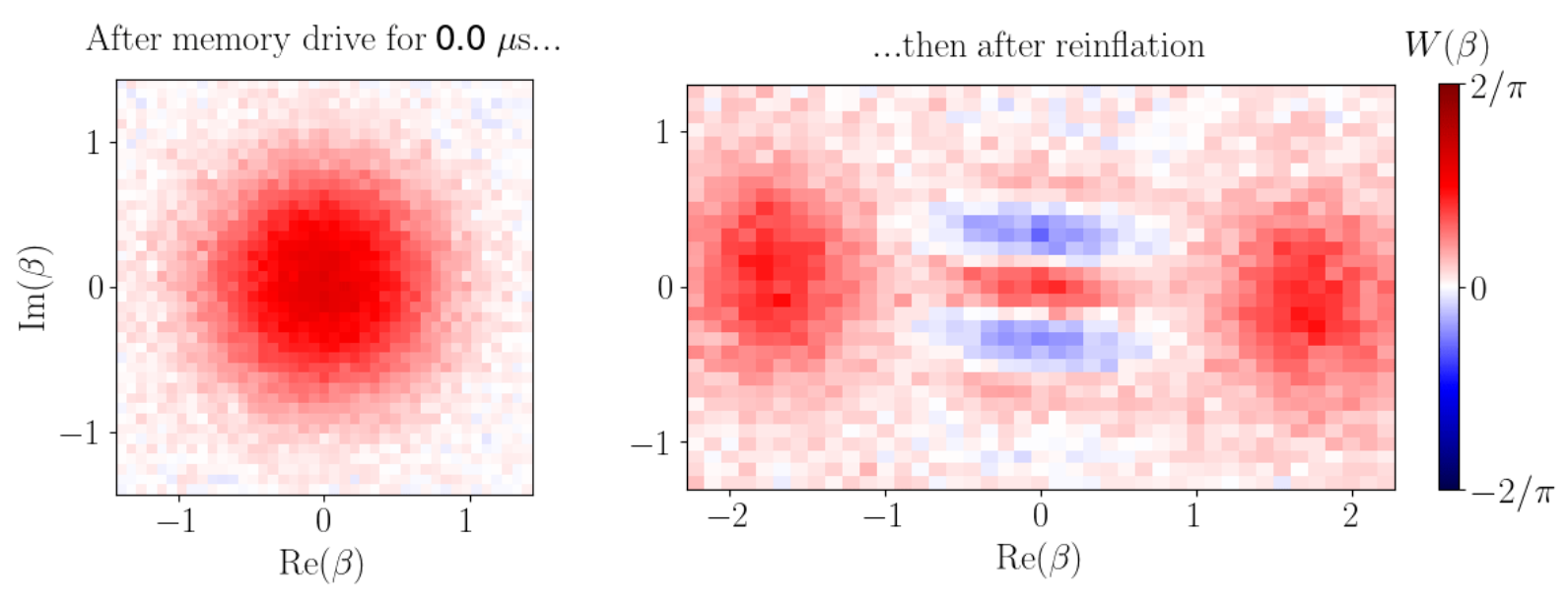}
  \caption{Video showing the measured Wigner functions of the memory for increasing driving time $t_\mathrm{rot}$ during the $\hat{Y}\left(\theta \right)$ gate sequence. See the video in this \href{https://arxiv.org/src/2403.07744v2/anc/Ytheta.gif}{ancillary file }.
  \label{vid:Ygate}
  }
  \end{video}

The corresponding gate infidelity can be explained by the duration of the displacement pulse $t_\mathrm{rot} > \kappa_2^{-1}$, during which errors have time to accumulate. These long times originate from the upper bound on $\epsilon_Y$ imposed by Quantum Zeno dynamics. The latter is notably much stricter than for the $\hat{Z}\left(\theta \right)$ gate demonstrated in~\cite{Marquet2023} for which one only requires  $\epsilon_\mathrm{Z} < 2 |\alpha|^2\kappa_2$ resulting in a $\hat{Z}$ gate as fast as $28~$ns. 

In contrast to most gates in bosonic qubits, the main source of errors during this gate is not single-photon loss but pure memory dephasing, characterized by a Kraus operator $\sqrt{\kappa_{\varphi}^m}\hat m^\dagger \hat m$ with $\kappa_{\varphi}^m = 80~$kHz $\approx 6 \kappa_1$. This term does not impact the memory when the cat state manifold is stabilized, owing to the two-photon dissipation $\hat{L}_2 = \sqrt{\kappa_2}\left(\hat{m}^2 - \alpha^2 \right)$ which suppresses the dephasing as long as $2 \alpha^2 \kappa_2 \gg \kappa_\varphi$ and for $\alpha\gtrsim 1$. In contrast, the manifold $\mathrm{span}\left(\ket{0}, \ket{1} \right)$ does not benefit from such protection. As a consequence, memory dephasing induces decoherence during the displacement at a rate $\kappa_{\varphi}^m$, resulting in decohered cat states after re-inflation. This issue could be mitigated by reducing the memory dephasing rate, in particular by working at a sweet spot of the autoparametric cat (see Appendix of~\cite{Marquet2023}).

Finally, we note that the holonomic gate presented in the main text spends a smaller amount of time in the manifold generated by $\{|0\rangle,|1\rangle\}$, resulting in a much higher gate fidelity.

\subsection{Holonomic $\hat{X} \left( \theta \right)$ gate}
\label{Sec:Holo_gate_appendix}

\subsubsection{Calibration of the gate}

The evolution of the joint memory-buffer system during the holonomic $\hat{X}$ gate can be simulated using the master equation

\begin{equation}
\begin{aligned}
\label{Eq:simu_Holo_X_gate}
    \frac{d\hat \rho}{dt} =& -\frac{i}{\hbar}[\hat{H}(t) , \hat \rho] + \kappa_1\mathcal{D}[\hat{m} ](\hat \rho)+ \kappa_{\varphi}^m\mathcal{D}[\hat{m}^\dagger \hat{m}](\hat \rho) \\
    &+ \kappa_b\mathcal{D}[\hat b](\hat \rho)+ \kappa_{\varphi}^b\mathcal{D}[\hat b^\dagger \hat b](\hat \rho), 
\end{aligned}
\end{equation}

\noindent with the time-dependent Hamiltonian expressed as

\begin{equation}
\begin{aligned}
\label{Eq:simu_Holo_X_gate_Hamiltonian}
    \frac{\hat{H}(t)}{\hbar} = \,  - &\frac{\chi_\mathrm{m,m}}{2} \hat{m}^{\dagger 2} \hat{m}^2 - \frac{\chi_\mathrm{b,b}}{2} \hat{b}^{\dagger 2} \hat{b}^2 - \chi_\mathrm{m,b} \hat{m}^\dagger \hat{m} ~\hat{b}^\dagger \hat{b}\\
    + \, & g_2\hat{m}^2\hat{b}^\dagger + g_2^*\hat{m}^{\dagger 2}\hat{b} + \epsilon_d(t)\hat{b}^\dagger + \epsilon_d^*(t) \hat{b}.
\end{aligned}
\end{equation}

\noindent Note that we do not adiabatically eliminate the buffer at this stage in order to simulate  the transient dynamics between the manifolds $\mathrm{span}\left(\ket{\alpha}, \ket{-\alpha} \right)$ and $\mathrm{span}\left(\ket{0}, \ket{1} \right)$ during which the condition $8 g_2 \alpha < \kappa_b$ is not necessarily verified. The values of the different parameters are summarized in Tab.~\ref{Tab:Device_parameters}. To optimize the gate fidelity, we perform numerical simulations for various shapes of the buffer drive $\epsilon_d(t)$ while enforcing the following constraints. Because the gate starts and finishes in the cat manifold, this gives a first constraint on $\epsilon_d(t)$

\begin{equation}
|\epsilon_d(0)| = |\epsilon_d(T_\mathrm{tot})| = |\epsilon_\alpha|,
\end{equation}

\noindent with $T_\mathrm{tot}$ the total duration of the gate and $|\epsilon_\alpha|$ the drive amplitude stabilizing a cat state of size $|\alpha|$. The second constraint comes from the necessity for the memory to transit through $\mathrm{span}\left(\ket{0}, \ket{1} \right)$ to realize the desired population transfer. As a consequence, there must exist a time $\tau$ during the gate where $\epsilon_d(\tau) = 0$. We choose $\tau = T_\mathrm{tot}/2$ due to the symmetry between the deflation and re-inflation steps of the gate. Furthermore, we impose that the evolution of $\epsilon_d(t)$ is symmetric around $\tau$. Finally, we choose a Gaussian edge shape with a standard deviation $\sigma$ for the evolution of the buffer drive for $0\leq t \leq \tau$. This choice is motivated by the observation that, while $\epsilon_d(t)$ needs to evolve on timescales longer than the inverse confinement rate $\kappa_\mathrm{conf}^{-1} = 1/(2 \kappa_2 |\alpha|^2)$, for the gate to remain adiabatic, faster evolutions are allowed for large $|\alpha|$ at the beginning and end of the gate. The buffer drive is then parameterized as 

\begin{equation*}
    \epsilon_d(t) = \frac{\epsilon_\alpha}{1 - e^{-\tau^2/(2 \sigma^2)}}\left(e^{-t^2/(2 \sigma^2)} - e^{-\tau^2/(2 \sigma^2)} \right) 
\end{equation*}

\noindent for $0 \leq t \leq \tau$, and 

\begin{equation*}
    \frac{\epsilon_\alpha e^{2i \theta}}{1 - e^{-\tau^2/(2 \sigma^2)}}\left(e^{-(2\tau - t)^2/(2 \sigma^2)} - e^{-\tau^2/(2 \sigma^2)} \right)
\end{equation*}

\noindent for $\tau \leq t \leq 2\tau$.

\begin{figure}[ht!]
\centering
\includegraphics[scale = 1.1]{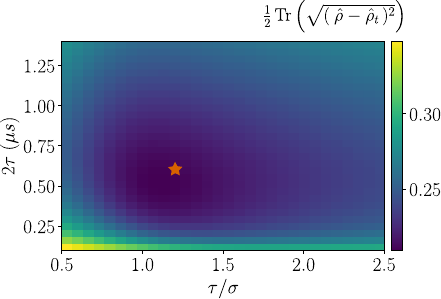}
\caption{Dependence on $\tau$ and $\tau / \sigma$ of the trace distance between the simulated density matrix $\hat{\rho}$ after running the gate pulse sequence while starting from $|\alpha\rangle$ and the targeted state $\hat{\rho}_t=\hat{X}\left(\pi /2 \right) \ket{\alpha} \bra{\alpha} \hat{X}^\dagger\left(\pi /2 \right)$ . In this simulation, $\alpha=2.5$. The star indicates the optimal parameters that were chosen in the experiments.} 
\label{fig:X_gate_opti}
\end{figure}

We optimize the parameters $\tau$ and $\sigma$ of this pulse by simulating the evolution of the memory according to Eq.~(\ref{Eq:simu_Holo_X_gate}). Starting from $\ket{\alpha}$, with $\alpha=2.5$, we compare the simulated density matrix $\hat{\rho}$ after the gate for $\theta = \pi/2$ with the target state $\hat{\rho}_{t} = \hat{X}\left(\pi /2 \right) \ket{\alpha} \bra{\alpha} \hat{X}^\dagger\left(\pi /2 \right)$. The metric used to compare these two density matrices is the trace distance~\cite{Nielsen2000}, defined as 

\begin{equation}
    T\left(\hat{\rho}, \hat{\rho}_{t} \right) = \frac{1}{2} \mathrm{Tr} \left(\sqrt{\left(\hat{\rho} - \hat{\rho}_{t} \right)^2} \right).\label{eq:tracedistance}
\end{equation}

\begin{figure}[ht!]
\centering
\includegraphics[scale = 1.1]{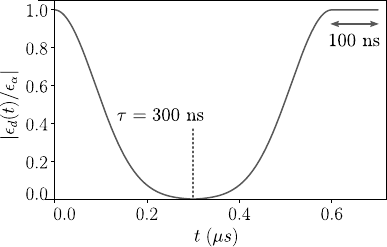}
\caption{Evolution of $|\epsilon_d(t)|$ for the chosen parameters $\tau$ and $\sigma$. A final stabilization time of $100$ ns is added after the gate to ensure the memory converges to the manifold $\mathrm{span}\left(\ket{\alpha}, \ket{-\alpha} \right)$.} 
\label{fig:X_gate_waveform}
\end{figure}

\noindent The optimization shown in Fig.~\ref{fig:X_gate_opti} provides the parameters $\tau = 300$ ns and $\sigma = \tau / 1.2 = 250$~ns. The corresponding waveform for $|\epsilon_d(t)|$ is presented in Fig.~\ref{fig:X_gate_waveform}.  

\subsubsection{Estimating the performance of the holonomic gate}

\begin{figure}[ht!]
\centering
\includegraphics[scale = 1.1]{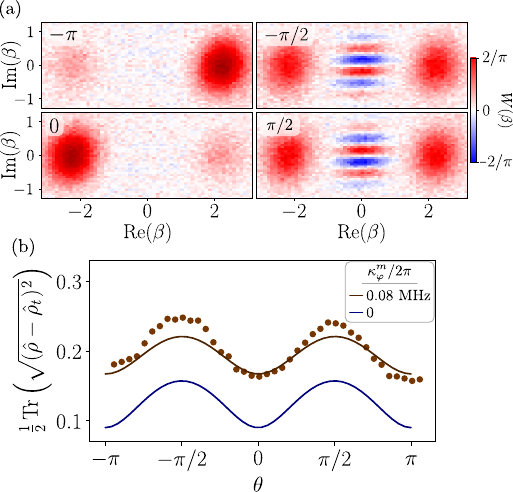}
\caption{(a) Measured Wigner function $W\left(\beta \right)$ after performing the gate $\hat{X}\left(\theta \right)$, starting from the coherent state $\ket{-\alpha}$. The angle $\theta$ corresponding to each Wigner function is indicated as an inset. Video~\ref{vid:Xgatedynamics} shows the same Wigner functions for every measured angle. (b) Dots: Trace distance between the density matrices $\hat{\rho}$ estimated from the Wigner functions of (a) and $\hat{\rho}_{t} = \hat{X}\left(\theta \right) \ket{-\alpha} \bra{-\alpha} \hat{X}^\dagger\left(\theta \right)$. Lines: Simulated trace distance for $\kappa_\varphi^\mathrm{m}/2\pi = 0.08~$MHz (brown line) and $\kappa_\varphi^\mathrm{m}/2\pi = 0~$MHz (blue line).} 
\label{fig:X_Holo_trace_distance}
\end{figure}

We study the performance of the holonomic gate by repeating the measurement shown in Fig.~\ref{fig:fig3} and Video~\ref{vid:Xdynamics} for varying values of $\theta$. Since the fidelity is quite low, we do not perform a full process tomography and simply compute how close the $\hat{X}\left(\theta \right)$ gate brings the coherent state $\ket{-\alpha}$ to its target $\hat{X}\left(\theta \right)|-\alpha\rangle$. The density matrix $\hat{\rho}$ is estimated from the measured Wigner function shown in Fig.~\ref{fig:X_Holo_trace_distance}a. using Maximum Likelihood Estimation (MLE). We impose that $\hat{\rho}$ is a density matrix in the Hilbert subspace $\mathrm{span}\left(\ket{\alpha}, \ket{-\alpha} \right)$, where $\alpha$ is deduced from the position of the coherent states in the measured Wigner function. 

\begin{video}
  \includegraphics[width=\columnwidth]{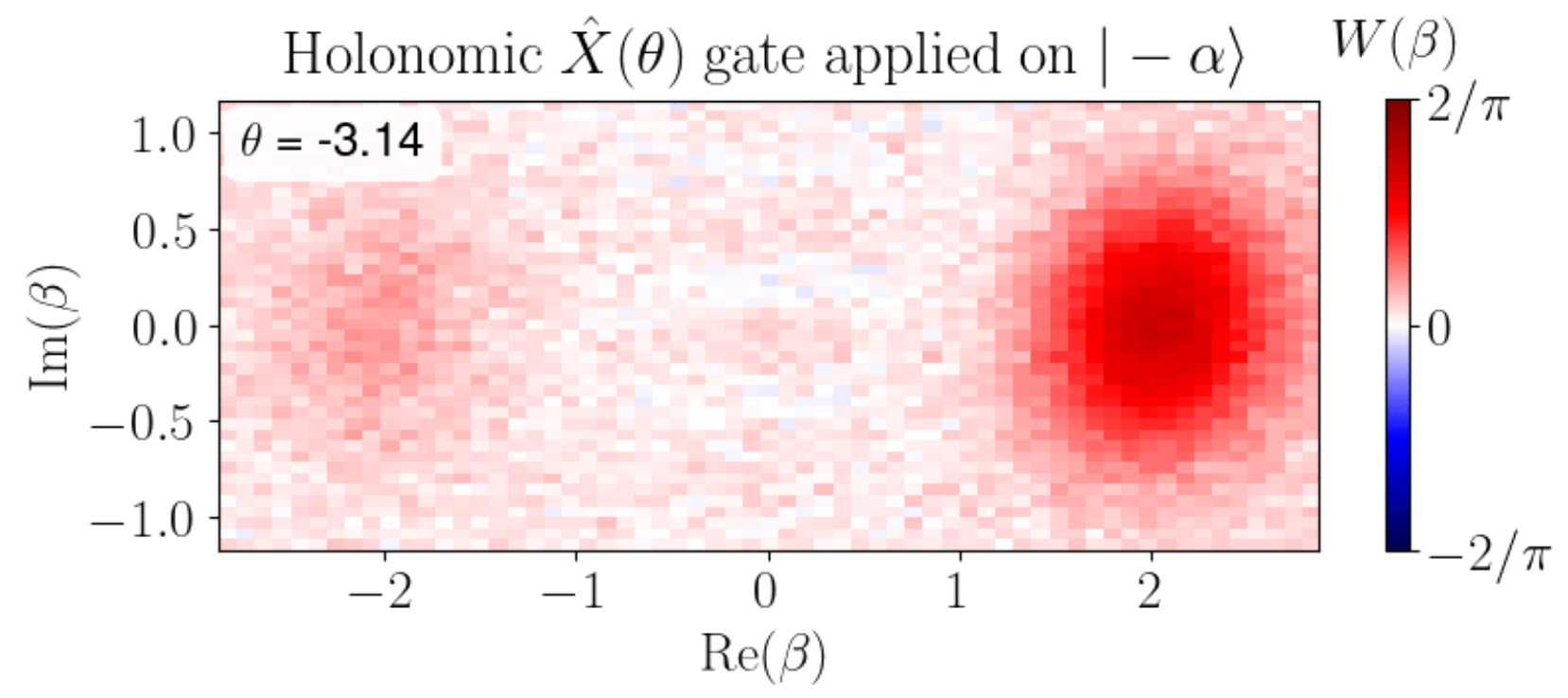}
  \caption{Measured Wigner function $W\left(\beta \right)$ after performing the gate $\hat{X}\left(\theta \right)$, starting from the coherent state $\ket{-\alpha}$ and for increasing angle $\theta$ with time. See the video in this \href{https://arxiv.org/src/2403.07744v2/anc/Xgate.gif}{ancillary file }.
  \label{vid:Xgatedynamics}
  }
  \end{video}

The dependence of the trace distance on $\theta$ is shown in Fig.~\ref{fig:X_Holo_trace_distance}b and Video~\ref{vid:Xgatedynamics}. A non-trivial evolution can be observed with a trace distance reaching $T\left( \hat{\rho}, \hat{\rho}_{t} \right) \approx 0.18$ for $\theta = \pm \pi,~0$ and up to $T\left( \hat{\rho}, \hat{\rho}_{t} \right) \approx 0.23$ for $\theta = \pm \pi/2$. We attribute the deterioration for $\theta = \pm \pi/2$ to phase-flip errors occurring during the inflation step of the pulse sequence and the final $100~$ns of stabilization (see Fig.~\ref{fig:X_gate_waveform}). Indeed, while for $\theta = \pm \pi,~0$ the memory ends up in the coherent states $\ket{\pm \alpha}$ which are unaffected by phase flip errors, the states $\ket{C_\alpha^{\pm i}}$ do not benefit from the same protection and lose their coherence at a rate $\Gamma_\mathrm{Z} = 2 \Bar{n} \kappa_1$. Note that while phase-flips also impact the memory during the deflation step of the gate, they act as a global increase of the trace distance (of a few \%) for all angle $\theta$ as the gate always starts from a common state $|-\alpha \rangle$.

In order to better understand the limitation on gate errors, we simulate the evolution of the trace distance according to the master equation Eq.~(\ref{Eq:simu_Holo_X_gate}), with the optimized buffer drive amplitude shown in Fig.~\ref{fig:X_gate_waveform}. Using the measured dephasing rate $\kappa_\varphi^\mathrm{m}/2\pi = 0.08~$MHz, we qualitatively reproduce the experimental results, although the trace distance is underestimated for $\theta = \pm \pi/2$ owing to measurement errors (Wigner tomography is less faithful when the state exhibit coherences). Additional simulations for $\kappa_\varphi^\mathrm{m} = 0$ indicate that reducing the dephasing rate could reduce the trace distance by $\approx 0.08$ using the same pulse parameters.

\subsubsection{Logical qubit state as a function of gate angle }

\begin{figure}[ht!]
\centering
\includegraphics[width=\columnwidth]{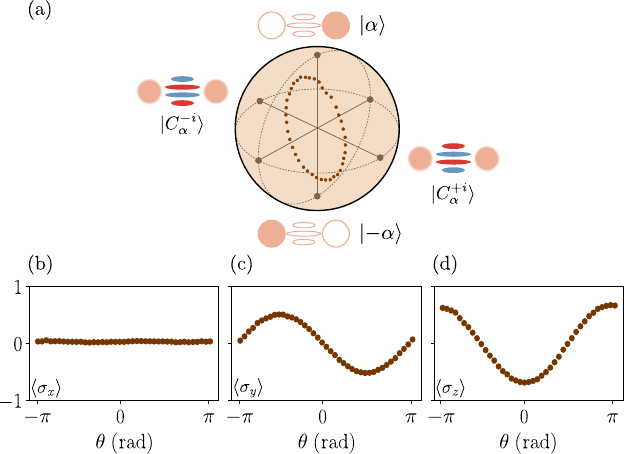}
\caption{(a) Bloch vectors of the cat qubit when applying a $\hat{X}\left( \theta\right)$ gate on the state $|-\alpha\rangle$ for various angles $\theta$. The Bloch vector is estimated from the Wigner tomography of Fig.~\ref{fig:X_Holo_trace_distance}(b) Mean value of $\hat{\sigma}_{x, L}$ as a function of the parameter $\theta$. (c) Mean value of $\hat{\sigma}_{y, L}$ as a function of the parameter $\theta$. (d) Mean value of $\hat{\sigma}_{z, L}$ as a function of the parameter $\theta$.} 
\label{fig:05_X_gate_fidelity_2D}
\end{figure}

We finally determine the Bloch vector of the memory in the logical cat qubit space. The logical Pauli operators $\hat{\sigma}_{x, L}$, $\hat{\sigma}_{y, L}$ and $\hat{\sigma}_{z, L}$ are defined on the basis of states $\{|\alpha\rangle,|-\alpha\rangle\}$. Fig.~\ref{fig:05_X_gate_fidelity_2D}(a) shows coherent state $|-\alpha\rangle$ rotated by a gate $\hat{X}(\theta)$ for various values of $\theta$. $\left< \hat{\sigma}_{x, L}(\theta) \right>$ shows no visible dependence on $\theta$ (Fig.~\ref{fig:05_X_gate_fidelity_2D}(b)). Its small mean value $\left< \hat{\sigma}_{x, L} \right> = 0.02$ can be attributed to single-photon loss. Indeed, during the time $\approx 200~$ns spent in the manifold $\mathrm{span} \left(\ket{0}, \ket{1} \right)$, single-photon loss induces the decay $\ket{1} \rightarrow \ket{0}$ which increases the population of even Fock states once the state is mapped back onto the cat manifold. The two remaining quadratures $\left< \hat{\sigma}_{y, L}(\theta) \right>$ and $\left< \hat{\sigma}_{z, L}(\theta) \right>$ exhibit the expected oscillations (Fig.~\ref{fig:05_X_gate_fidelity_2D}(c-d)), with a phase shift of $\pi/2$ between the 2 curves. A small phase shift of $0.12~$rad is observed, particularly noticeable in Fig.~\ref{fig:05_X_gate_fidelity_2D}(d) where $\left< \hat{\sigma}_{z, L}(\theta) \right>$ is not symmetric around the axis $\theta = 0$. This can be explained with an additional detuning $\Delta_m \hat{a}^\dagger \hat{a}$, $\Delta_m/2\pi = 90~$kHz to the Hamiltonian of Eq.~(\ref{Eq:simu_Holo_X_gate_Hamiltonian}). This induces a rotation in the memory phase space when in the manifold $\mathrm{span} \left(\ket{0}, \ket{1} \right)$, with no impact when the 2-photon dissipation stabilizes $\mathrm{span} \left(\ket{\alpha}, \ket{-\alpha} \right)$ at a rate $\kappa_2 \gg \Delta_m$.

\begin{video}
  \includegraphics[width=\columnwidth]{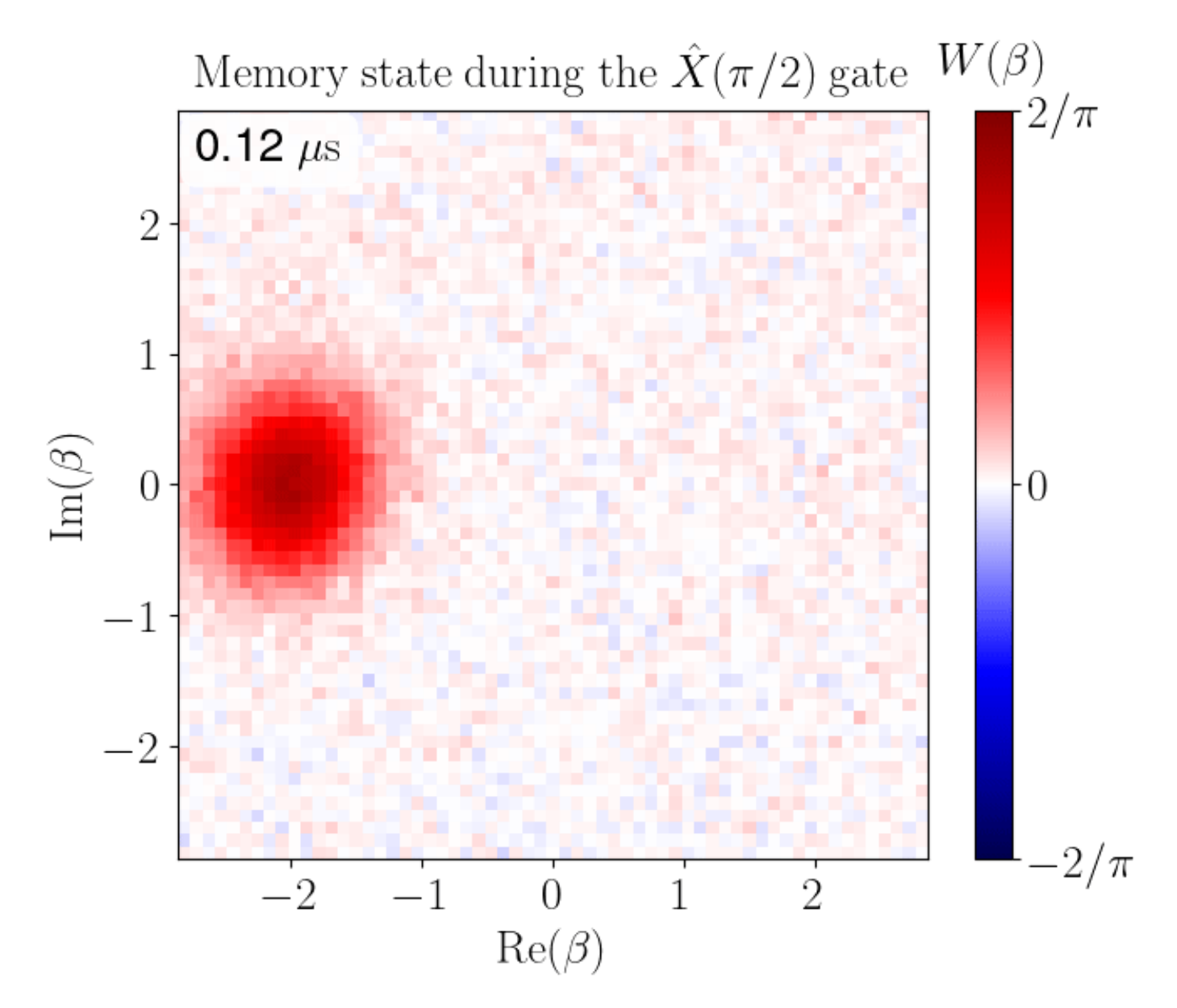}
  \caption{Measured evolution of the memory Wigner function, starting from the coherent state $\ket{-2.27}$, during the $\hat{X}\left(\pi/2 \right)$ gate. This video extends Fig.~\ref{fig:fig3}. See the video in this \href{https://arxiv.org/src/2403.07744v2/anc/Xpi2dynamics.gif}{ancillary file}.
  \label{vid:Xdynamics}
  }
  \end{video}

\subsection{Comparison to state-of-the-art}

In this section, we compare the gate fidelities we obtain to the state-of-the-art. In this work and in Ref.~\cite{Marquet2023}, we demonstrate gate errors $\epsilon_{Z_{\pi/2}}=2~\%$ (excluding state preparation and measurement (SPAM) errors) and estimate $\epsilon_{X_{\pi/2}}=23~\%$ (including about 2~\% of SPAM errors).  These are the smallest infidelities for dissipation stabilized cat qubits. 

The smallest infidelities to date for Kerr cat qubits are obtained in Ref.~\cite{Hajr2024}. They demonstrate $\epsilon_{Z_{\pi/2}}=8~\%$ (including about $7~\%$ of SPAM errors) and $\epsilon_{X_{\pi/2}}=9~\%$ (including about $7~\%$ of SPAM errors).

Finally, while it is difficult to compare with our protected two-component cat qubits, to our knowledge the smallest gate infidelities for unprotected 4 component cat states are obtained in Ref.~\cite{Heeres2017} with about $1.5~\%$ infidelity.

\section{Timescale associated to squeezing}
\label{Sec:squeezing}

Turning off the buffer drive results in a squeezing drive on the memory mode at short times, as shown in Fig.~\ref{fig:fig4}. In this section, we estimate the rate of the effective squeezing drive and the timescale over which it is effective. The squeezing dynamics can be captured qualitatively when neglecting the self-Kerr effect, dephasing, memory loss and detuning between $2\omega_m$ and $\omega_b$. 

The dynamics of the system composed of memory and the undriven buffer is ruled by the master equation, 
\begin{align}
    \frac{d\rho}{dt} &= -i[\hat{H},\rho] + \kappa_b \mathcal{D}[\hat{b}](\rho) \\
    \hat{H} &=  g_2 \hat{m}^2 \hat{b}^\dagger + g_2^* \hat{m}^{\dagger 2} \hat{b}.
\end{align}
We take $g_2 >0$ and assume the memory is initially in a cat state $|C_\alpha^+\rangle$ with $\alpha>0$. We consider sufficiently short times for which the memory mode remains in a superposition of two states with negligible overlap. It is then possible to use a time-dependent shifted Fock basis~\cite{Chamberland2022}, centered on the time-dependent amplitude $m(t)$ of the cat states, given by the transformation $ \hat{m} \rightarrow \hat{\sigma}_Z\otimes(\tilde{m}+m(t))$. 
Here, $\hat{\sigma}_Z$ is the Pauli Z operator in the qubit basis $\{(|C_{m(t)}^+\rangle+|C_{m(t)}^-\rangle)/\sqrt{2},(|C_{m(t)}^+\rangle-|C_{m(t)}^-\rangle)/\sqrt{2}\}$, while $\tilde{m}$ is the annihilation operator of the so-called gauge mode.  We also make a displacement on the buffer mode by the transformation $\hat{b} \rightarrow \tilde b + \gamma(t)$. The master equation becomes 
\begin{align}
    \frac{d\rho_\mathrm{sq}}{dt} &= -i[\tilde{H},\rho_\mathrm{sq}] + \kappa_b \mathcal{D}[\tilde b](\rho_\mathrm{sq}) \notag \\
    \tilde{H} &=  g_2 \tilde m^2 \tilde b^\dagger + 2 g_2 m(t) \tilde m \tilde b^\dagger + \text{h.c} \notag \\
    &+ g_2 \gamma^* (t) \tilde m^2 + \text{h.c}  \\
    & -i\frac{d\gamma}{dt}\tilde b^\dagger + g_2 m^2(t)\tilde b^\dagger - i\frac{\kappa_b}{2}\gamma(t) \tilde b^\dagger + \text{h.c}  \\
    & -i \frac{d m}{dt} \tilde m^\dagger + 2 g_2 m^*(t) \gamma(t) \tilde m^\dagger + \text{h.c} 
\end{align}

The amplitudes $m(t)$ and $\gamma(t)$ are chosen such as they cancel the linear drive terms on the buffer and memory modes, i.e they cancel the last two lines of the above equation. The amplitudes $m(t)$ and $\gamma(t)$ hence satisfy the differential equations

\begin{align}
    \frac{d m}{dt} &= -i 2 g_2 \gamma m^* \notag \\
    \frac{d\gamma}{dt} &= -\frac{\kappa_b}{2}\gamma - i g_2 m^2. \label{eq:amp_diff}
\end{align}
From these equations and the initial conditions $\gamma(0)=0$ and $m(0) = \alpha$, one deduces that $r \in\mathbb{R}$, $m \in\mathbb{R}$ and $\gamma\in i\mathbb{R}$, i.e the cat-like state remains aligned along the $x$-axis in phase-space ($\beta$ real), while the buffer mode moves along the $y$-axis ($\beta$ purely imaginary).

The integral forms of the solutions read 
\begin{align}
    m(t) &= \alpha e^{-r(t)}, ~~ r(t) = i 2 g_2\int_0^t \gamma(\tau)d\tau  \label{eq:m_amp} \\
    \gamma(t) &= -i g_2 \int_0^{t}d\tau e^{-\frac{\kappa_b}{2}(t-\tau)}m^2(\tau) \label{eq:b_amp}, 
\end{align}
where we made use of $\gamma(0)=0$ and $m(0) = \alpha$. As we are interested in describing the dynamics of the system at short times, we make a Taylor expansion of the factor $r(t) = \lambda_2 t^2 + \lambda_3 t^3 + O(t^4)$. Here, we used $r(0) = \frac{dr}{dt}(0) = 0$. Inserting this expression into the above equations leads to 
\begin{equation}r(t) = g_2^2 \alpha^2 t^2 (1- \frac{\kappa_b}{6} t) + O(t^4).\end{equation} 

Let us show that the parameter $r(t)$ corresponds to the squeezing parameter.
Having chosen the displacement amplitudes satisfying Eq.~(\ref{eq:amp_diff}) leads to the master equation 
\begin{align}
    \frac{d\tilde\rho}{dt} &= -i[\tilde{H},\tilde\rho] + \kappa_b \mathcal{D}[\tilde b](\tilde\rho) \notag \\
    \tilde{H} &=  \big( g_2 \tilde m^2 \tilde b^\dagger + 2 g_2 m(t) \tilde m \tilde b^\dagger + \text{h.c} \big) + \big(\frac{i}{2} \frac{dr}{dt} \tilde m^2 + \text{h.c}\big).
\end{align}

The last term corresponds to a squeezing drive. It can be eliminated by moving into the squeezed frame defined by the unitary transformation $\rho_\text{sq} = U_\text{sq}^\dagger \tilde{\rho} U_\text{sq}$ with $U_\text{sq}= \exp\big(\frac{r}{2}(\tilde m^2 - \tilde{m}^{\dagger^2}) \big)$. 
In the displaced and squeezed frame, the master equation reads 
\begin{align}
    \frac{d\rho_\mathrm{sq}}{dt} &= -i[H_\mathrm{sq},\rho_\mathrm{sq}] + \kappa_b \mathcal{D}[\tilde b](\rho_\mathrm{sq}) \notag \\
    H_\mathrm{sq} &=   g_2 (\cosh(r) \tilde m - \sinh(r)\tilde m^\dagger )^2 \tilde b^\dagger \notag \\
    &+ 2 g_2 m(t) (\cosh(r) \tilde m - \sinh(r)\tilde m^\dagger ) \tilde b^\dagger + \text{h.c} .
\end{align}
It is worth noting that while this master equation describes the system in a frame that is displaced and squeezed, the two-photon exchange term between the buffer and the memory remains active.
The squeezed mode $\tilde m $ looses photons through its coupling to the buffer. It also thermalizes due to the coupling of the form $\tilde m^\dagger\tilde{b}^\dagger$ at a rate bounded by $\kappa_b/2$.

\renewcommand{\arraystretch}{1.3}

\begin{table*}[!th]
\centering
\caption{\textbf{Estimated parameters of the device and the associated measurement method.}}
\begin{tabular}{c|c|c}
\hline
\hline
\multicolumn{1}{c}{\textbf{Parameter}} & \multicolumn{1}{c}{\textbf{Value}} & \multicolumn{1}{c}{\textbf{Method of determination}}\\
\hline
2 photon dissipation flux $\phi_\mathrm{on}$ & 0.312 $\phi_0$  & Memory and buffer spectroscopy \\

Tomography flux $\phi_\mathrm{off}$ & 0.168 $\phi_0$  & Optimization of the memory displacements $\hat{\mathcal{D}}\left( \beta \right)$\\

Sweet spot $\phi_\mathrm{ext}^\mathrm{(sweet)}$ & 0.4 $\phi_0$  & Memory and buffer spectroscopy\\

Memory frequency $\omega_m\left(\phi_\mathrm{on} \right)/2\pi$ & $3.948$~GHz & Memory and buffer spectroscopy\\

Buffer frequency $\omega_b\left(\phi_\mathrm{on} \right)/2\pi$ & $7.896$~GHz & Memory and buffer spectroscopy\\

Transmon frequency $\omega_\mathrm{q}/2\pi$ & $5.387$~GHz & Ramsey interferometry at $\phi_\mathrm{off}$\\

Readout resonator frequency $\omega_\mathrm{r}/2\pi$ & $6.967$~GHz & Reflection measurement\\

Effective inductive energy $\bar{E}_J\left(\phi_\mathrm{on} \right)/\hbar$ & $228$~GHz & Memory and buffer spectroscopy\\

Effective inductive energy $\bar{E}_W\left(\phi_\mathrm{on} \right)/\hbar$ & $51$~GHz & Memory and buffer spectroscopy\\

Effective inductive energy $\bar{E}_J\left(\phi_\mathrm{off} \right)/\hbar$ & $242$~GHz & Memory and buffer spectroscopy\\

Effective inductive energy $\bar{E}_W\left(\phi_\mathrm{off} \right)/\hbar$ & $97$~GHz & Memory and buffer spectroscopy\\

Memory participation ratio $\varphi_{\mathrm{ZPF}, m}$ & 0.0305 & Memory and buffer spectroscopy\\

Buffer participation ratio $\varphi_{\mathrm{ZPF}, b}$ & 0.0648 & Memory and buffer spectroscopy\\

Predicted two-to-one coupling rate $g_2/2\pi$ & $6.2~\mathrm{MHz}$ & Memory and buffer spectroscopy\\

\hline

Memory self-Kerr $\chi_\mathrm{m, m}$ & $0.220$~MHz & Coherent state deformation\\

Buffer self-Kerr $\chi_\mathrm{b, b}$ & $10$~MHz & Estimated using Appendix B of~\cite{Marquet2023}\\

Transmon to memory cross-Kerr $\chi_\mathrm{q, m}$ & $0.170$~MHz & Ramsey interferometry with a displaced resonator\\

Buffer to memory cross-Kerr $\chi_\mathrm{b, m}$ & $1.6$~MHz & Estimated using Appendix B of~\cite{Marquet2023}\\

Transmon to readout cross-Kerr $\chi_\mathrm{q, r}$ & $3.5$~MHz & Reflection measurement\\

Two-to-one coupling rate $g_2/2\pi$ & $6 \pm 0.5~$MHz  & Fidelity of logical $\hat{Z}$ rotations\\

\hline

Memory single photon loss $\kappa_1/2\pi$ & $14$~kHz & Decay $\ket{1} \rightarrow \ket{0}$\\

Memory effective 2 photons loss $\kappa_2/2\pi$ & $2.16$~MHz & Decay $\ket{C_\alpha^+} \rightarrow \ket{0}$ or $\ket{C_\alpha^-} \rightarrow \ket{1}$ \\

Memory dephasing rate $\kappa_\varphi^m/2\pi$ & $0.08$~MHz & Ramsey interferometry Zeno blocked on $\mathrm{span}\left(\ket{0}, \ket{1} \right)$\\

Buffer single photon loss $\kappa_b/2\pi$ & $40$~MHz & Reflection measurement\\

Readout resonator linewidth $\kappa_r/2\pi$ & $1.8$MHz & Reflection measurement\\

Transmon $T_1$ & $18~\mu \mathrm{s}$ & Standard decay measurement \\

Transmon $T_2$ & $15~\mu \mathrm{s}$ & Ramsey interferometry\\

Memory thermal population $n_\mathrm{th, m}$ & $0.011$ photon & Ramsey interferometry\\

Transmon thermal population $n_\mathrm{th, q}$ & $0.015$ photon & Standard transmon readout\\

\hline
\hline
\end{tabular}
\label{Tab:Device_parameters}
\end{table*}

\end{document}